\documentclass[twocolumn,showpacs,preprintnumbers,amsmath,amssymb,floatfix]{revtex4-1}

\usepackage{graphicx}
\usepackage{longtable}
\usepackage{dcolumn}
\usepackage{bm}
\usepackage{fancybox}

\newcommand{\bfr}{{\bf r}}
\newcommand{\hbfr}{\hat{\bf r}}

\newcommand{\bfT}{{\bf T}}
\newcommand{\bfG}{{\bf G}}
\newcommand{\bfR}{{\bf R}}

\newcommand{\refeq}[1]{Eq.~(\ref{#1})}

\newcommand{\YY}{{\cal Y}}
\newcommand{\GG}{{\cal G}}

\newcommand{\ooplus}{\oplus}
\newcommand{\oominus}{\ominus}

\def\tphi{{\tilde{\phi}}}
\def\calR{{\cal A}}

\def\phidot{\dot{\phi}}

\def\H0{H^0}

\newcommand{\req}[1]{\mbox{Eq.~\!(\ref{#1})}}
\newcommand{\refsec}[1]{\mbox{Sec.~\!\ref{#1}}}

\def\connect#1{\leavevmode{\setbox1=\hbox{#1}\copy1%
\raise .2\ht1 \vbox{\moveleft \wd1\vbox{\hrule width \wd1 height .5pt depth 0pt}}%
}}

\def\smh{smHankel}
\def\smhs{smHankels}

\def\ftn[#1]{\rlap{\footnotemark[#1]}}

\def\EMAXm{ E^{\rm rmesh}_{\rm MAX} }

\def\RSM{R_{\rm SM}}

\def\RGSa{R_{{\rm G},a}}
\def\pakl{p_{akl}}
\def\PakL{P_{akL}}
\def\wPakL{\widetilde{P}_{akL}}
\def\CakL{C_{akL}}
\def\CiakL{C^i_{akL}}

\def\EMAXm{ E^{\rm rmesh}_{\rm MAX} }

\def\nc{n^{\rm c}}
\def\nzc{n^{\rm Zc}}
\def\nzcv{n^{\rm Zcv}}
\def\barnzcv{\bar{n}^{\rm Zcv}}
\def\MM{{\cal M}}
\def\RR{w}
\def\inta{\int_{|\bfr|\leq R_a}\!\!\!\!\!\!\!\!\!\!\!\!}
\def\intaa{\int_{|\bfr|\leq R_a}}
\def\intad{\int_{|\bfr'|\leq R_a}\!\!\!\!\!\!\!\!\!\!\!\!}

\def\rhoij{\rho_{ij}}
\def\ekcore{E_{\rm k}^{\rm core}}
\def\ek{E_{\rm k}}
\def\ehf{E_{\rm Harris}}
\def\nin{n^{\rm in}}
\def\nout{n^{\rm out}}
\def\Vin{V}

\def\philo{{\phi}^{\rm Lo}_{al}}

\begin{document}

\title{Formulation of the augmented plane-wave and muffin-tin orbital
method}
\author{Takao Kotani}
\affiliation{Department of applied mathematics and physics, Tottori university, Tottori 680-8552, Japan}
\author{Hiori Kino}
\affiliation{National Institute for materials science, Sengen 1-2-1, Tsukuba, Ibaraki 305-0047, Japan.}
\author{Hisazumu Akai}
\affiliation{Institute of solid state Physics, Tokyo university, Kashiwa 277-8581, Japan}
\date{\today}

\begin{abstract}
A mixed basis all-electron full-potential method, 
which uses two kinds of augmented waves, the augmented plane waves  
and the muffin-tin orbitals simultaneously, in addition to the
local orbitals, was proposed by Kotani and van Schilfgaarde in
Phys. Rev. B81, 125117(2010). We named it the PMT method.
In this paper, this mixed basis method is reformulated on the basis of a new
formalism named as the 3-component formalism, which is a
mathematically transparent version of the additive
augmentation originally due to Soler and Williams in Phys. Rev. B47, 6784(1993). 
Atomic forces are easily derived systematically. 
We discuss some problems in the mixed basis method and ways to
manage them. In addition, we 
compare the method with the PAW method on the same footing.
This PMT method is the basis for our new development of the quasiparticle
self-consistent $GW$ method in J.Phys.Soc.Jpn 83, 094711(2014).
\end{abstract}
\maketitle

\section{introduction}
In the first-principles electronic-structure calculations based on the
density functional theory in the LDA/GGA
(local density approximation/generalized gradient approximation),
a key element is the one-body problem solver, which 
should have efficiency, accuracy and robustness.
As such solvers, the linearized augmented plane wave (LAPW) method and the 
linearized muffin-tin orbital (LMTO) method were proposed by Andersen
in 1975 \cite{Andersen75}, followed by many improvements
and extensions \cite{rmartinbook,Singhbook,bluegel31,lmfchap,PAW,PhysRevB.43.6388}.
Today LAPW and LMTO has developed to be full-potential methods, 
which we treat in this paper.
In these methods, wavefunctions are represented by superpositions of
augmented waves. The LAPW uses the augmented plane waves (APWs) made 
of plane waves (PWs) as envelope functions; 
the LMTO uses the muffin-tin orbitals (MTOs) made of the atom-centered 
Hankel functions. Corresponding to these envelope functions, the APWs fit to
the extended nature of wavefunctions; 
in contrast, the MTOs to the localized nature of them.
However, wavefunctions in real materials should have both the natures.

This fact is reflected as shortcomings in these methods.
In the case of the LAPW, it requires so many bases 
in order to represent sharp structures of wavefunctions 
just outside of muffin-tins. For example, 3$d$ orbitals of transition
metals are the typical cases. Most of all the PWs used in the LAPW method 
are consumed only to reproduce the sharp structures.
On the other hand, the LMTO is problematic in representing 
the extended nature of wavefunctions. For example, we sometimes need to 
put empty spheres between atoms. In addition, it is not 
simple to enlarge basis set systematically 
in order to check numerical convergence.

To overcome these shortcomings, 
Kotani and van Schilfgaarde introduced a new linearized method 
named as the APW and MTO method (the PMT method) \cite{pmt1}, which is an all-electron (AE) 
full-potential mixed basis method using APWs and MTOs simultaneously.
Because these two kinds of basis are complementary, 
we can overcome these shortcomings. Within our knowledge, 
no other mixed basis methods have used different kinds 
of augmented waves simultaneously in the full-potential methods.

A minimum description on the formalism of the PMT method is given in Ref.\cite{pmt1}, 
which is based on Ref.\cite{lmfchap} for a LMTO method  
by Methfessel et al. However, the formalism was not very transparent, 
mainly because it was not derived from the explicit total energy minimization.
This makes theoretical treatment of the PMT method somehow complicated.
For example, it resulted in a complicated logic to derive atomic forces 
in Refs.\cite{molforce,lmfchap}. 
It was not easy to compare the PMT method with the projector augmented
wave (PAW) methods \cite{PAW,kresse99} on the same footing.
Thus we should give a simple and clear formalism to the PMT method 
for its further developments rather than that in Refs.\cite{pmt1,lmfchap}.

In this paper, we introduce a formalism, named as the 3-component
formalism, which is a mathematically transparent generalization of the additive
augmentation given by Soler and Williams \cite{soler89,soler90,soler93} 
(See discussion in Sec.VII in Ref.\cite{PAW}).
We give a formalism of the PMT method based on the 3-component formalism.
In the PAW method \cite{PAW}, the total energy is minimized as a
functional of pseudo wavefunctions. In the 3-component formalism,
the minimization is formulated as for the wavefunctions represented in the 
{\it 3-component space} defined in \refsec{sec:formalism} under some constraints.
This is somehow general in the sense that it allows to use any kinds of basis 
(need not to be given by projectors); thus it is
suitable to formulate mixed basis methods such as the PMT.
Results of the PMT method applied to diatomic molecules from H$_2$ through Kr$_2$ 
are already given in Ref.\onlinecite{kotani_linearized_2013}. 
Considering the fact that the PMT method (even the LMTO itself) is already pretty good to
describe solids \cite{pmt1,lmfchap,kotani07a,mark06adeq},
the PMT method can be a candidate to perform full-potential
calculations for molecules and solids in an unified manner, 
more efficiently than LAPW.

Note that we had already implemented 
the quasiparticle self-consistent $GW$ (QSGW) method
\cite{Faleev04,mark06qsgw,kotani07a} in the PMT method
\cite{kotani_quasiparticle_2014}; this allows us to apply the
QSGW method to wide range of materials without empty spheres, 
with not being bothered with the difficulty of setting parameters of
MTOs. We have applied the QSGW method to cases
\cite{han_quasiparticle_2014,nagara_band_2014}.
This kind of method was reffered to by
Kimes, Katak and Kresse \cite{klimes_predictive_2014}, who claimed that 
accurare and efficient $GW$ calculations should be based on a method
using both kinds of bases (localized and extended bases in space) simultaneously.

In Sec.\ref{sec:formalism}, we will give the 3-component formalism.
Functional relations of physical quantities become transparent in the formalism.
In Sec.\ref{sec:pmtmethod}, we give the formulation of the PMT method
based on the 3-component formalism. Then we discuss problems
in the PMT method and ways to overcome them, giving a comparison with the PAW method.
Derivation of atomic forces becomes straightforward as given in Appendix
without any confusion that were discussed in Ref.\cite{soler93}.

\section{3-components formalism}
\label{sec:formalism}
We assume periodic boundary condition where
real space (or unit cell) is specified by $\Omega$.
$\Omega$ is divided into the muffin-tin (MT) regions and the interstitial region.
The MTs are located at $\bfR_a$ with radius $R_a$, where 
$a$ is the index specifying a MT within $\Omega$.
$L\equiv(l,m)$ is the angular momentum index.
We use units, $\hbar=1$, electron mass $m_e$, and electron charge $e$. 
The spin index is not shown for simplicity.

Here we give the 3-component formalism as a general
scheme for the augmented-wave methods, which include any kinds of
augmented waves including the local orbitals \cite{PhysRevB.43.6388}.

\subsection{the 3-component space}
\label{sec:3compo}
Any augmented basis $F_{i}(\bfr)$ consists of three kinds of components, 
where $i$ is the index specifying basis function. $F_{i}(\bfr)$
consists of the following three components:
\begin{itemize}
\item[(0)] 
the smooth part (= envelope function) $F_{0i}(\bfr)$
\item[(1)]
the true parts $F_{1i,a}(\bfr)$ defined in MTs $|\bfr| \leq R_a$.
\item[(2)]
the counter parts $F_{2i,a}(\bfr)$ defined in MTs $|\bfr| \leq R_a$ 
(canceling the smooth parts within MTs).
\end{itemize}
We call $F_{0i}(\bfr)$, $F_{01,a}(\bfr)$, and $F_{02,a}(\bfr)$ as the
0th, 1st, and 2nd components of $F_{i}(\bfr)$, respectively.
The $F_{0i}(\bfr)$ should be an analytic function in space or a linear
combination of analytic functions.
In the PMT method, $F_{0i}(\bfr)$ are the PWs or the Bloch sums of the
Hankel functions; exactly speaking, we use atom-centered 
smooth Hankel functions (\smhs) instead of the conventional Hankel functions, 
so as to avoid divergence at its center 
\cite{lmfchap,Bott98} (See \req{eq:defh0} and around). 
$F_{1i,a}(\bfr)$ and $F_{2i,a}(\bfr)$ are defined only at
$|\bfr| \leq R_a$ (in cases below, we sometimes take these are zero at $|\bfr|>R_a$).
In the sense that $F_{0i}(\bfr)$ is analytic and 1st and 2nd components 
are given on a dense radial mesh, a basis $F_{i}(\bfr)$ is specified
without any numerical inaccuracy.

$F_{i}(\bfr)$ is a member in the {\it 3-component space}, which
is defined as a direct sum of linear spaces
corresponding to the components (0), (1) and (2). Thus $F_i$ can be expressed as
$F_i=\{F_{0i}(\bfr),\{F_{1i,a}(\bfr)\},\{F_{2i,a}(\bfr)\} \}$
(curly bracket mean a set), where $F_{1i}\equiv \{F_{1i,a}(\bfr)\}$
means a set as for the MT index $a$, $F_{2i}$ as well. However, in the followings, we use a
little different expression instead:
\begin{eqnarray}
F_i(\bfr) = F_{0i}(\bfr) \ooplus \{ F_{1i,a}(\bfr)\} \oominus \{F_{2i,a}(\bfr)\}\nonumber\\
= F_{0i}(\bfr) \ooplus F_{1i}(\bfr)\oominus F_{2i}(\bfr).
\label{eq:basis}
\end{eqnarray}
This makes following expressions easy to read without any difference
in their meanings. Symbols $\ooplus$ and $\oominus$ mean nothing more
than separators.
We call a member in the 3-component space as a 3-component function in
the followings.
Wavefunctions are also given as 3-component functions.
With the coefficients $\{\alpha_{p}^i\}$, wavefunctions can be written as
\begin{eqnarray}
\psi_p(\bfr) = \sum_i \alpha_p^i F_i(\bfr),
\label{eq:eig}
\end{eqnarray}
where linear combinations are taken for each components.
We represent electron density and so on as a 3-component function as well.

Note that the 3-component space is a mathematical
construction, a model space: we have to specify how to map a 3-component 
function to a function in real space.
For this purpose, we define $\calR$-mapping (augmentation mapping)
from a 3-component function to a function in real space;
\begin{eqnarray}
\calR[\psi_p(\bfr)]\!\equiv\! \psi_{0p}(\bfr)
\!+\! \sum_a \psi_{1p,a}(\bfr\!-\!\bfR_a)\!-\!\sum_a \psi_{2p,a}(\bfr\!-\!\bfR_a). \nonumber \\
\label{eq:calRF}
\end{eqnarray}
This is nothing but a conventional augmentation 
where physically meaningful wavefunctions $\psi_{p}(\bfr)$ 
should satisfy following conditions (A) and (B);
\begin{itemize}
\item[(A)]
Within MTs ($|\bfr|<R_a$), $\psi_{2p,a}(\bfr)=\psi_{0p}(\bfr+\bfR_a)$.
\item[(B)]
At MT boundaries ($|\bfr|=R_a$), $\psi_{1p,a}(\bfr)$ and $\psi_{2p,a}(\bfr)$
should have the same value and slope.
\end{itemize}
If (A) is satisfied, the contribution from $\psi_{0p}$ within MTs 
perfectly cancels those of $\psi_{2p,a}$ in \req{eq:calRF}. 
The total energy in the DFT is given as a functional
of eigenfunctions as $E[\{\psi_{p}(\bfr)\}]$, where
$\{\psi_{p}(\bfr)\}$ are for occupied states.
Our problem is to minimize this under the constraint of
orthogonality of $\psi_{p}(\bfr)$ with conditions 
(A) and (B) on $\{\psi_{p}(\bfr)\}$.
Local orbitals \cite{PhysRevB.43.6388} is also treated as 3-component
functions whose 0th and 2nd components are zero overall.

In the conventional LAPW (e.g. See \cite{bluegel31,Singhbook}),
(A) and (B) are very accurately satisfied. The 2nd component
almost completely satisfy (A) with the use of spherical Bessel functions.
The 1st component are given up to very high $l$ ($\gtrsim 8$). 
Thus the LAPW can be quite accurate. However, it can be 
expensive (we also have null-vector problem. See \refsec{sec:problems}.).

Thus Soler and Williams \cite{soler89} introduced additive augmentation:
to make calculations efficient, we use condition (A') as a 
relaxed version of condition (A),
\begin{itemize}
\item[(A')]
Within MTs ($|\bfr|\leq R_a$), 
$\psi_{2p,a}(\bfr) \approx \psi_{0p}(\bfr+\bfR_a)$.
\end{itemize}
Then we expect high-energy (high frequency) contributions
of eigenfunctions not included in the 1st and 2nd components 
are accounted for by the 0th component. In practice, we can use low $l$ 
cutoff $\lesssim 4$ for both of 1st and 2nd components. 
A LAPW package HiLAPW, developed by Oguchi et al \cite{PhysRevB.54.1159},
implemented a procedure to evaluate physical quantities from 
the basis given by \req{eq:calRF} with the condition (A').

However, it is complicated to evaluate all quasilocal 
products such as the density and kinetic-energy density
from $\calR[F^*_i(\bfr)]\calR[F_j(\bfr')]$,
since it contains cross terms which connect different components.
Thus Soler and Williams \cite{soler89} 
gave a prescription to avoid the evaluation of the cross terms.
With $\calR$-mapping applied not to wavefunctions
but to products of them as in \refsec{aug3},
we have {separable form} of the 
total energy and all other physical quantities 
(no cross terms between components). This is based on the fact that
the total energy in the {separable form} should 
agree with the true total energy only when (A) and (B) are satisfied. 
As we see in the followings, it is a good approximation to use (A') instead of (A). 

Above two important concepts, 
the additive augmentation and the separable form, 
were used in both of LMTO and PAW \cite{lmfchap,PAW,kresse99}.
They were originally introduced in Ref.\cite{soler89}.

Let us consider how to determine $F_{1i,a},F_{2i,a}$ for
a given $F_{0i}$. As for $F_{2i,a}$, (A') means that 
$F_{0i}$ should be reproduced well within MTs.
Generally speaking, $F_{2i,a}(\bfr)$ can be represented as
\begin{eqnarray}
&&F_{2i,a}(\bfr) \equiv \sum_{k,L} \CiakL P_{akL}(\bfr), \label{f2}
\end{eqnarray}
where $k$ is index for radial degree of freedom. 
We introduce truncation parameters
$k_{{\rm max},a}$ and $l_{{\rm max},a}$; we assume 
sum in \req{f2} is taken for $k\le k_{{\rm max},a}$ and $l \le l_{{\rm max},a}$;
when $k_{{\rm max},a}$ and $l_{{\rm max},a}$ becomes infinite, we assume
condition (A) is satisfied. Even when these truncation
parameters are finite, $F_{2i,a}$ should reproduce
low energy (low frequency) parts of $F_{0i}$ well. 
The functions $\{P_{akL}(\bfr)\}$ can be rather general; 
as explained in \refsec{sec:pmtmethod}
the central parts of \smh\ is treated as it is
(in other words, treated as a member of $\{P_{akL}(\bfr)\}$ \cite{privatemark1}).
$F_{1i,a}(\bfr)$ is given from \refeq{f2}
with a replacement of ${P}_{akL}(\bfr)$ with $\widetilde{P}_{akL}(\bfr)$.
Here $\widetilde{P}_{akL}(\bfr)$ is a linear combination of
partial waves so as to have the same value and slope with
${P}_{akL}(\bfr)$ at $|\bfr|=R_a$. With this replacement, we have 
\begin{eqnarray}
F_{1i,a}(\bfr) = \sum_{k,L} C^i_{akL} \wPakL(\bfr). \label{f1}
\end{eqnarray}

\subsection{augmentation for product of 3-component functions}
\label{aug3} Let us give a prescription to evaluate physical quantities
for wavefunctions satisfying conditions (A') and (B).  First, we define
diagonal product of 3-component functions as
\begin{widetext}
\begin{eqnarray}
F^*_i(\bfr)F_{j}(\bfr') &\equiv& F^*_{0i}(\bfr)F_{0j}(\bfr')
\ooplus  \{ F^*_{1i,a}(\bfr) F_{1j,a}(\bfr')\}
\oominus \{ F^*_{2i,a}(\bfr) F_{2j,a}(\bfr')\}, \label{fnproduct}
\end{eqnarray}
where we have no cross terms between different components.
We apply $\calR$-mapping in \req{eq:calRF} to this product as
\begin{eqnarray}
&&\calR[F^*_i(\bfr)F_{j}(\bfr')] = \nonumber \\ 
&&F^*_{0i}(\bfr)F_{0j}(\bfr')
+ \sum_a F^*_{1i,a}(\bfr\!-\!\bfR_a) F_{1j,a}(\bfr'\!-\!\bfR_a)
- \sum_a F^*_{2i,a}(\bfr\!-\!\bfR_a)F_{2j,a}(\bfr'\!-\!\bfR_a). \label{fnmap}
\end{eqnarray}
\end{widetext}
We will use $\calR[F^*_i(\bfr)F_{j}(\bfr')]$ to evaluate quasilocal 
products when (A') is satisfied.
Since any one-body quantities such as the inner product, electron density,
current and so on, are quasilocal, we can evaluate these from 
$\calR[\psi^*_p(\bfr)\psi_{p'}(\bfr')]$.
Generally speaking, we can evaluate matrix elements of a
quasilocal operator $X(\bfr,\bfr')$ in real space
from 3-component wavefunctions $\psi_p(\bfr)$ in separable form as 
\begin{eqnarray}
\langle \psi_p|X|\psi_{p'} \rangle=\int d^3r d^3r' X(\bfr,\bfr') 
\calR[\psi^*_p(\bfr)\psi_{p'}(\bfr')].
\end{eqnarray}
We can read this as a transformation of $X$ to the corresponding 
operator in the 3-component space.

Based on the above prescription, we can define the inner product 
$\langle \psi_p |\psi_{p'} \rangle$ as $\langle \psi_p |\psi_{p'}
\rangle=\sum_{i,j}\alpha_{p}^{i*}\alpha_{p'}^{j}O_{ij}$.
Here the overlap matrix  $O_{ij}$ is:
\begin{widetext}
\begin{eqnarray}
O_{ij} &\equiv& \langle F_i|F_{j} \rangle \equiv \int_\Omega d^3r 
  \calR[F^*_i(\bfr)F_j(\bfr)] \nonumber \\
&=&\int_\Omega d^3r  F^*_{0i}(\bfr)F_{0j}(\bfr)
  + \sum_a \inta d^3r  F^*_{1i,a}(\bfr)F_{1j,a}(\bfr) 
  - \sum_a \inta d^3r  F^*_{2i,a}(\bfr)F_{2j,a}(\bfr).  \label{eq:norm} 
\end{eqnarray}
This can read as a definition of the inner product in the 3-component space.
For a given finite basis set, we can expect that $O_{ij}$ should be
positive definite as long as truncation parameters are large enough.
The kinetic energy is given 
from $\rho_{ij}=\sum_{p}^{\rm occ.} \alpha^{i*}_p \alpha^j_p$ (occ. means
the sum for occupied states) as $\ek=\sum_{i,j} \rho_{ij} T_{ij}$.
Here the kinetic-energy matrix $T_{ij}$ is given as
\begin{eqnarray}
&&T_{ij}\equiv \frac{\langle \nabla F_i| \nabla F_{j} \rangle}{2m_e}
 \equiv \frac{1}{2m_e} \int_\Omega d^3r \left(\nabla_\bfr \nabla_{\bfr'}
 \calR[F^*_i(\bfr)F_j(\bfr')]\right)_{\bfr=\bfr'} 
=\frac{1}{2m_e} \int_\Omega d^3r 
\calR[\nabla F^*_i(\bfr) \nabla F_j(\bfr)] \nonumber \\
&&= \int_\Omega d^3r \frac{\nabla F^*_{0i}(\bfr) \nabla F_{0j}(\bfr)}{2m_e} 
+ \sum_a \inta d^3r \frac{\nabla F^*_{1i,a}(\bfr) \nabla F_{1j,a}(\bfr)}{2m_e}
- \sum_a \inta d^3r \frac{\nabla F^*_{2i,a}(\bfr) \nabla F_{2j,a}(\bfr)}{2m_e}.
\label{eq:kin}
\end{eqnarray}
Partial integration gives $T_{ij}= \langle F_i| \frac{-\nabla^2 }{2m_e}|F_{j} \rangle$, 
since $F_{1i,a}$ and $F_{2i,a}$ have the same value and slope at the MT boundaries.
This kinetic energy operator is interpreted as $T=\frac{-\nabla^2 }{2m_e} \oplus
\{\frac{-\nabla^2 }{2m_e} \} \ominus  \{\frac{-\nabla^2 }{2m_e} \}$ in the 3-component space.

One-body problem for a given one-particle potential $V(\bfr)$ in real
space is translated into a problem in the 3-component space for the
Hamiltonian $H=T+V$ under the condition (A) or (A'), where $V=V_0 \oplus
\{V_{1,a}\} \ominus \{V_{2,a}\}$.  Here $V_0(\bfr)=V(\bfr)$, and
$V_{1,a}(\bfr)=V_{2,a}(\bfr)=V(\bfr+\bfR_a)$ within MTs at
$\bfR_a$. However, we can add any extra potential $\Delta \bar{V}$
simultaneously to both of $V_0$ and $V_{2,a}$ if (A) is completely
satisfied.

We have an error because we use \req{fnmap} instead of \req{eq:calRF}:
high energy contributions contained in the 0th components are not
exactly evaluated. However, the error can be small enough to be
neglected as discussed in Appendix \ref{sec:zeroonetwo}.  This error is
also related to a question, how to choose the optimum $\Delta \bar{V}$
so as to minimize the error.  In fact, the success of the PAW \cite{PAW}
is dependent on the choice of $\Delta \bar{V}$ as seen in
\refsec{sec:comparison}.

The valence electron density $n$ as the 3-component function is given by
\begin{eqnarray}
n &=&n_0\ooplus n_1 \oominus n_2 = n_0 \ooplus \{n_{1,a}\} \oominus \{n_{2,a}\} =
\sum_{ij} \rhoij F_i^* F_j = \sum_{ij}
 \rhoij F^*_{0i}(\bfr)F_{0j}(\bfr) \nonumber \\
&&\ooplus  \{ \sum_{ij} \rhoij F^*_{1i,a}(\bfr)F_{1j,a}(\bfr) \}
\oominus   \{ \sum_{ij} \rhoij F^*_{2i,a}(\bfr)F_{2j,a}(\bfr) \}. \label{eq:n}
\end{eqnarray}
\end{widetext}
We can calculate the Coulomb interaction from $\calR[n]$. 
However, to reduce the computational effort, 
we will also make the Coulomb interaction into the separable form as
seen in \refsec{sec:coulomb}, with the help of multipole technique
due to Weinert \cite{weinert81}.
In \refsec{sec:multi} and \refsec{sec:frozencore}, we give some
preparations to define the Coulomb interaction in \refsec{sec:coulomb}.

The total energy should be given as a functional of eigenfunctions in
the first-principle calculations, not just as a functional of
coefficients $\{\alpha^j_p\}$.  This is important in some cases. For
example, it is necessary to know how the change in the basis set affects
the total energy when we calculate atomic forces. These are related to
the so-called Pulay terms \cite{pulay69}.

\subsection{multipole transformation}
\label{sec:multi} In order to define Coulomb interaction in
Sec.\ref{sec:coulomb}, we introduce the multipole transformation
($\MM$-transformation) for the the 3-component functions.  This
corresponds to the compensation charges in Ref.\cite{PAW}.

Before defining the $\MM$-transformation, we define the gaussian
projection $\GG_a \left[f(\bfr)\right]$ as follows.  The projection
$\GG_a \left[f(\bfr)\right]$ is defined for the function $f(\bfr)$ for
$|\bfr|\leq R_a$ as
\begin{eqnarray}
&&\GG_a \left[f(\bfr)\right]
=\sum_L Q_{aL}[f] G_{aL}(\bfr), \label{eq:gdef} \\
&&G_{aL}(\bfr)= \frac{1}{N_{aL}} \exp\left(-\left(\frac{r}{\RGSa}\right)^2\right)
Y_L(\hat{\bfr}),\label{eq:gl}
\end{eqnarray}
where $Q_{aL}[f]=\intaa \YY_L(\bfr) f(\bfr) d^3r$
gives the $L$-th multipole moment of $f(\bfr)$.
Here $\YY_L(\bfr) \equiv r^l Y_L(\hat{\bfr})$.
$Y_L(\hat{\bfr})$ is the real spherical harmonics, 
where $\hat{\bfr}$ is the normalized $\bfr$. 
$N_{aL}$ is a normalization factor
so that $G_{aL}(\bfr)$ has a normalized multipole moment.
$\RGSa$ in \req{eq:gl} is chosen small enough so that $G_{aL}(\bfr)$ is
negligible for $|\bfr|\geq R_a$ (See Eq.(25) in
Ref.\cite{lmfchap}). This $\GG_a \left[f(\bfr)\right]$ is a
superposition of gaussians $G_{aL}(\bfr)$ with keeping the multipole
moments of $f(\bfr)$.  We can take rather small $\RGSa$ without loss of
numerical accuracy; it is possible to take a limit $\RGSa \to 0$ because
quantities involved in $G_{aL}(\bfr)$ are evaluated analytically or
numerically accurately on a dense radial mesh.

We now define $\MM$-transformation for
3-component density $n= n_0 \ooplus n_1 \oominus n_2$ as
\begin{widetext}
\begin{eqnarray}
&&\MM[n]=  n_0(\bfr) \nonumber \\
&&+ \sum_{a,\bfT,L} Q_{aL}[n_{1,a}\!-\!n_{2,a}]G_{aL}(\bfr-\bfR_a-\bfT) 
\ooplus n_{1} \oominus 
\{n_{2,a}(\bfr) + \sum_L Q_{aL}[n_{1,a}\!-\!n_{2,a}]G_{aL}(\bfr) \}. \label{eq:multi}
\end{eqnarray}
\end{widetext}
Thus $\MM[n]$ adds the same gaussians to both of the 0th and 2nd
components.  $\bfT$ is the translational vectors of $\Omega$.  With this
transformation, the multipole moments of the 1st and 2nd components become
the same.  Note that the $\MM$-transformation is not a physically
meaningful transformation because $\calR[\MM[n]]=\calR[n]$.  With this
transformation, interstitial electrostatic potential calculated from the
0th component of \req{eq:multi} should be the same as that calculated
from $\calR[n]$.

\subsection{Coulomb interaction}
\label{sec:coulomb} In principle, we can define the Coulomb interaction
between $n(\bfr)=n_0\ooplus n_1 \oominus n_2$ and $m(\bfr)=m_0 \ooplus
m_1 \oominus m_2$ from the densities $\calR[n]$ and $\calR[m]$.  We can
use $\calR[\bar{n}]$ instead of $\calR[n]$ where $\bar{n}=\MM[n]$
satisfies $\calR[\bar{n}]=\calR[n]$, and $\calR[\bar{m}]$ as well.  Thus
the Coulomb interaction $\left(n|v|m\right)_{\rm original}$ is given as
\begin{widetext}
\begin{eqnarray}
\left(n|v|m\right)_{\rm original}=\sum_{\bfT} \int_\Omega d^3r d^3r' 
\calR[\bar{n}(\bfr)] v(\bfr-\bfr'+\bfT) \calR[\bar{m}(\bfr')].
\label{eq:coulomb0}
\end{eqnarray}
Here $v(\bfr)=e^2/|\bfr|$; $\sum_{\bfT}$ implicitly includes the
division by number of cells. Equation~(\ref{eq:coulomb0}) can not be
easily evaluated because $v(\bfr-\bfr'+\bfT)$ contains the cross terms
which connect the 0th component with other components.

Thus we use an approximation 
\begin{eqnarray}
\left(n|v|m\right)\equiv\MM[n] \cdot v \cdot \MM[m] = \bar{n} \cdot v \cdot \bar{m}, \label{eq:coulomb}
\end{eqnarray}
instead of $\req{eq:coulomb0}$, where dot operator for the 3-component
functions is given as
\begin{eqnarray}
&&\bar{n} \cdot v \cdot \bar{m} \equiv
\bar{n}_0 \bullet v \bullet \bar{m}_0 + \bar{n}_{1} \circ v \circ
\bar{m}_{1} - \bar{n}_{2} \circ v \circ \bar{m}_{2}\\
&&\bar{n}_0 \bullet v \bullet \bar{m}_0 \equiv \sum_{\bfT} \int_\Omega d^3r d^3r' \bar{n}_0(\bfr)v(\bfr-\bfr'+\bfT) \bar{m}_0(\bfr'), \label{eq:n0vn0}\\
&&\bar{n}_1 \circ v \circ \bar{m}_1 \equiv 
\sum_a \inta d^3r \intad d^3r' \bar{n}_{1,a}(\bfr)v(\bfr-\bfr') \bar{m}_{1,a}(\bfr'), \label{eq:n1vn1}\\
&&\bar{n}_2 \circ v \circ \bar{m}_2 \equiv \sum_a \inta d^3r \intad d^3r' \bar{n}_{2,a}(\bfr)v(\bfr-\bfr') \bar{m}_{2,a}(\bfr'). \label{eq:n2vn2}
\end{eqnarray}
Note that $X \bullet Y$ means integral over $\Omega$, whereas $X \circ
Y$ means integrals within MTs.

Let us evaluate the difference between \req{eq:coulomb0} and \req{eq:coulomb}.
This can be evaluated with the identity in Appendix \ref{sec:zeroonetwo} as
\begin{eqnarray}
&&\left(n|v|m\right)_{\rm original}-\left(n|v|m\right)= 
\sum_a \inta d^3r \intad d^3r'\Big(
\left( \bar{n}_0(\bfr) \!-\! \bar{n}_2(\bfr) \right) 
v(\bfr\!-\!\bfr') \left( \bar{m}_1(\bfr')\!-\!\bar{m}_2(\bfr') \right)  \nonumber \\
&&+\left( \bar{n}_1(\bfr)\!-\!\bar{n}_2(\bfr) \right) 
v(\bfr\!-\!\bfr') \left( \bar{m}_0(\bfr')\!-\!\bar{m}_2(\bfr') \right)\Big).
\label{eq:coulombdiff}
\end{eqnarray}
This is essentially the same with Eq.(13) in Ref.\cite{kresse99}.  In
\req{eq:coulombdiff}, the difference consists of contributions from MT
sites without terms connecting different MT sites. This is because
$\bar{n}_{1,a}(\bfr)$ and $\bar{n}_{2,a}(\bfr)$ have the same multipole
moments.  Since $\bar{n}_0(\bfr') - \bar{n}_2(\bfr)$ is high-$l$ or
highly oscillating part, and $\bar{n}_{1,a}(\bfr)-\bar{n}_{2,a}(\bfr)$
has zero multipole moments and zero at MT boundaries, we expect that the
separable form of \req{eq:coulomb} should be justified. We can check
this with changing the truncation parameters $l_{{\rm max},a}$ and 
$k_{{\rm max},a}$.

From \req{eq:coulomb}, we have the expression of the Coulomb
interaction as
\begin{eqnarray}
\left(F^*_i F_j|v|F^*_{i'}F_{j'}\right) = \MM[F^*_iF_j] \cdot v \cdot \MM[F^*_{i'}F_{j'}].
\label{eq:ffvff}
\end{eqnarray}
Here $F^*_i F_j$ is the diagonal product defined in \req{fnproduct} at
$\bfr=\bfr'$.  In calculations such as arising in the $GW$
approximations \cite{kotani07a}, we have to evaluate this as accurately
as possible so that the exchange-pair cancellation is kept well.

\subsection{Frozen core approximation}
\label{sec:frozencore} We often need to treat spillout of the core
density outside of MTs explicitly. Then we use the frozen core
approximation; the charge density due to the cores are evaluated by a
superposition of rigid cores as follows \cite{lmfchap}.

First, we perform a self-consistent atomic calculation under the
spherical approximation without a spin polarization to obtain its core
density $\nc_a(\bfr)$. Then we make a fitted density $\nc_{{\rm
sH},a}(\bfr)$ given by a linear combination of several \smh\ functions
so that $\nc_{{\rm sH},a}(\bfr)$ reproduces $\nc_{a}(\bfr)$ for
$|\bfr|>R_a$ within a numerical accuracy. Since $\nc_{{\rm H},a}(\bfr)$
are analytic and smooth at their centers, we can treat them numerically
accurately (we can use other kinds of analytic functions such as
gaussians instead of \smh\ functions).
  
Thus we have the expression of all the core electron density with adding
contribution from nucleus $-Z_a\delta(\bfr)$:
\begin{eqnarray}
\nzc = \sum_{a,\bfT} \nc_{{\rm sH},a}(\bfr-\bfR_a-\bfT) \ooplus
\{\nc_{a}(\bfr)-Z_a\delta(\bfr) \} \oominus \{\nc_{{\rm sH},a}(\bfr) \}.
\label{eq:nzc}
\end{eqnarray}
Applying the $\MM$-transformation to $\nzc$ gives
\begin{eqnarray}
\MM[\nzc]
&=& \sum_{a,\bfT} 
\left( \nc_{{\rm sH},a}(\bfr-\bfR_a-\bfT)+\sum_L Q_{aL}^{\rm Zc} G_{aL}(\bfr-\bfR_a-\bfT) \right)\ooplus
\{\nc_{a}(\bfr)-Z_a\delta(\bfr)\} \nonumber \\
&& \oominus \{ \nc_{{\rm sH},a}(\bfr)+\sum_L Q_{aL}^{\rm Zc}G_{aL}(\bfr)\},\label{eq:nc} \\
Q^{\rm Zc}_{aL} &=& Q_{aL}[\nzc_1-\nzc_2] =
Q_{aL}[\nc_{a}(\bfr)-Z_a\delta(\bfr)-\nc_{{\rm sH},a}(\bfr)] \label{eq:qalnzc}.
\end{eqnarray}

\subsection{total energy in density functional}
\label{sec:total} Let us give the total energy $E_{\rm total}$ for the
DFT, and construct the Kohn-Sham equation from it. With the kinetic
energy $\ek=\frac{1}{2m_e} \sum_{ij} \rhoij \langle \nabla F_i| \nabla
F_{j} \rangle$ from Eq.(\ref{eq:kin}), the total energy is given as:
\begin{eqnarray}
E_{\rm total}=\ekcore+ \ek+E_{\rm es}+E_{\rm xc},
\label{eq:etot}
\end{eqnarray} 
where $\ekcore$ is the kinetic energy of frozen cores as a constant.
$E_{\rm es}$ and $E_{\rm xc}$ are electrostatic and exchange-correlation
energies, respectively.
$E_{\rm es}$ is given as the electrostatic energy for the
total density $\nzcv =\nzc +n $, which are given in
Eqs(\ref{eq:n},\ref{eq:nzc}). 

Based on the definition \req{eq:ffvff}, we have
\begin{eqnarray}
E_{\rm es}= \frac{1}{2} (\nzcv|v|\nzcv)=\frac{1}{2} \MM[\nzcv] \cdot v \cdot \MM[\nzcv],
\label{eq:es}
\end{eqnarray}
where a constant due to the self-interaction of nucleus is implicitly
removed. 
Components of $\barnzcv(\bfr)=\MM[\nzcv]$ are given as
\begin{eqnarray}
&&\barnzcv_0(\bfr)= \nzc_{0}(\bfr)
+ \sum_{a,L,\bfT} 
(Q_{aL}^{\rm Zc}+Q^{\rm v}_{aL})G_{aL}(\bfr-\bfR_a-\bfT)+ n_0(\bfr),
\label{eq:barn0zcv}\\
&&\barnzcv_{1,a}(\bfr)=\nzc_{1,a}(\bfr) +n_{1,a}(\bfr), \label{eq:barn1zcv}\\
&&\barnzcv_{2,a}(\bfr)=\nzc_{2,a}(\bfr) 
+\sum_L (Q_{aL}^{\rm Zc}+Q^{\rm v}_{aL})G_{aL}(\bfr)+n_{2,a}(\bfr),\label{eq:barn2zcv}
\end{eqnarray}
where $Q^{\rm v}_{aL}=Q_{aL}[n_{1,a}-n_{2,a}]$.  We expand
$F^*_{0i}(\bfr)F_{0j}(\bfr)$ of $n_0$ in $\{e^{i \bfG \bfr}\}$ (to
obtain coefficients, $F^*_{0i}(\bfr)F_{0j}(\bfr)$ is tabulated on a
real-space mesh, then it is Fourier transformed).  The cutoff on $\bfG$
is specified by $\EMAXm$.  Then the 0th components in \req{eq:barn0zcv}
is represented by sum of analytic functions. Thus we can finally
calculate $\frac{1}{2} \barnzcv_0(\bfr) \bullet v \bullet
\barnzcv_0(\bfr)$ in $E_{\rm es}$.  Terms between gaussians located at
different MT sites are evaluated with the Ewald sum treatment.  The
terms related to MTs in $E_{\rm es}$ is $ \frac{1}{2} \barnzcv_1 \circ
\RR \circ \barnzcv_1 - \frac{1}{2} \barnzcv_2 \circ \RR \circ
\barnzcv_2$, which is calculated on a radial mesh accurately.

The exchange correlation term can be defined as
\begin{eqnarray}
E_{\rm xc}[\nzcv] = E_{\rm xc}[\nzcv_0] 
+  \sum_a E_{\rm xc}[\nzcv_{1,a}]  
-  \sum_a E_{\rm xc}[\nzcv_{2,a}].
\label{eq:exc}
\end{eqnarray}
The functional derivatives of $E_{\rm xc}[\nzcv]$
with respect to each component of $\nzcv$ gives
\begin{eqnarray}
v^{\rm xc} = v^{\rm xc}_0(\bfr)
\ooplus  \{v^{\rm xc}_{1,a}(\bfr)  \}
\oominus \{v^{\rm xc}_{2,a}(\bfr)  \}.
\label{eq:vxc}
\end{eqnarray}

To determine the ground state, 
$E_{\rm total}$ should be minimized under the orthogonality of eigenfunctions with
the constraint (A') and (B).
This ends up with 
$\delta \psi^*_p \cdot (H - \epsilon_p) \cdot \psi_p =0$ for
the variation $\delta \psi^*_p$ which satisfy (A') and (B). 
Here the operator $H=T+V$ is given as
\begin{eqnarray}
&&T=\frac{-\nabla^2 }{2m_e} \oplus \left\{\frac{-\nabla^2 }{2m_e}\right\} 
\ominus \left\{\frac{-\nabla^2 }{2m_e}\right\} \\
&&V= 
\barnzcv_0 \bullet v \bullet + v^{\rm xc}_0
\ooplus
\left\{ \sum_L
{\cal Q}^{\rm v}_{aL} \YY_L(\bfr)
+ \barnzcv_{1,a} \circ \RR \circ + v^{\rm xc}_{1,a} 
\right\} 
\oominus
\left\{ \sum_L
{\cal Q}^{\rm v}_{aL} \YY_L(\bfr)
+ \barnzcv_{2,a} \circ \RR \circ + v^{\rm xc}_{2,a} 
\right\}, \label{eq:v} \\
&&{\cal Q}^{\rm v}_{aL}\equiv \frac{\partial E_{\rm es}}{\partial {Q}^{\rm v}_{aL}} =
\barnzcv_0 \bullet v \bullet G_{aL}(\bfr'-\bfR_a)
-\barnzcv_{2,a} \circ \RR \circ G_{aL}(\bfr'), \label{eq:calqdef}
\end{eqnarray}
where $\bar{n}_0 \bullet v \bullet$ means an integral on a variable, resulting a function of $\bfr$.

When a basis set $\{F_j(\bfr)\}$ satisfying (A') and (B) are fixed, we
just need to consider variation with respect to $\alpha_{p}^{i*}$ in
\req{eq:eig}. Then we have
\begin{eqnarray}
\sum_j (H_{ij} -\epsilon_p O_{ij}) \alpha_p^j =0,
\label{eq:eigenp}
\end{eqnarray}
\end{widetext}
where $H_{ij}= \langle F_i| H |F_{j} \rangle = 
\langle F_i| \frac{-\Delta}{2m} + V |F_{j} \rangle =T_{ij}+V_{ij}$.
$V_{ij}=\langle F_i|V|F_{j} \rangle=V \cdot F^*_i F_j$. 
Then the total energy minimization results in the eigenvalue problem.
The matrix elements $O_{ij},T_{ij}$
and $V_{ij}$ are given in Appendix \ref{onsitematrix}.

The formula to evaluate atomic forces are given in Appendix
\ref{sec:force}. It is directly evaluated from the variation on the
total energy. This procedure is considerably simplified than that given
in Refs.\cite{lmfchap,molforce}. 


\section{PMT method}
\label{sec:pmtmethod} Let us give the PMT method based on the
3-component formalism in \refsec{sec:formalism}. Based on it, we need to
specify a basis set $\{F_i\}$.  In the PMT, $\{F_i\}$ is classified into
three kinds of subsets as follows:
\begin{itemize}
\item[(a)] APW. We augment the PW in the manner as will be shown later.

\item[(b)] MTO. We augment the atom-centered \smh\ functions. 

\item[(c)] Local orbital (Lo) \cite{lo}.
   We use this to represent some degree of freedom in MTs,
   such as semicore states. The envelope function of Lo is zero overall.

\end{itemize}

The \smh\ function, as the envelop function of MTO, is first introduced
by Methfessel \cite{lmfchap,Bott98}. The spherical \smh\ function
$h_0(\bfr)$ (for $l=0$) is defined by the Helmholtz equation with a
gaussian source term $g_0(\bfr) = C \exp(-r^2/\RSM^2)$ (see Eq.(5) in
Ref.\cite{lmfchap}) instead of $\delta$-function;
\begin{equation}
(\nabla^2+\epsilon)h_0(\bfr) = -4 \pi g_0(\bfr),
\label{eq:defh0}
\end{equation}
where $C=1/(\sqrt{\pi} \RSM)^3$ is the normalization constant.
$\epsilon=-\kappa^2$ is the negative energy to specify the asymptotic
damping behavior of $h_0(\bfr)$.  At the limit $\RSM \to 0$ where
$g_0(\bfr)$ becomes $\delta$-function (as a point charge), $h_0(\bfr)$
becomes to the Hankel function $h_0(\bfr)=\exp(-\kappa r)/r$.  Since the
source term is smeared with the radius $\RSM$, we have no divergent
behavior at $r=0$ anymore; the \smh\ bends over at $\sim\RSM$ (See Fig.1
in Ref.\cite{lmfchap}).  From $h_0(\bfr)$, we can make
$h_L(\bfr)\equiv\YY_L(-\nabla) h_0(\bfr)$ for any $L$.
(recall $\YY_L(\bfr)=r^l Y_L(\hat{\bfr})$).
$\YY_L(-\nabla)$ means to substitute $\bfr$ in $\YY_L(\bfr)$
with $-\nabla$. See Ref.\cite{Bott98} for details.

For the augmentation of the PW, that is, to determine the 2nd component
from PW as 0th component, we expand the PW within the MTs into the
Laguerre polynomial \cite{pmt1}.  Any function $f(\bfr)$ (PW in this
case) is expanded within a MT $|\bfr-\bfR_a|\leq R_a$ as
\begin{eqnarray}
&&f(\bfr)= \sum_{k,l} \CakL[f] \PakL(\bfr-\bfR_a), \label{onecenter1}\\
&&\PakL(\bfr)=\pakl(r) Y_L(\hbfr) \label{onecenter2},
\end{eqnarray}
where $k=0,1,2,...$ denotes the order of a polynomials $\pakl(r)$.  In
the case that $f(\bfr)$ is a PW, the coefficients for the function
$\CakL[f]$ are given analytically \cite{Bott98}.

When we use \smh\ centered at $\bfR_a$ as an envelope function
$f(\bfr)$, 
we have head part, which is $f(\bfr)=h_L(\bfr-\bfR_a)$ for
$|\bfr-\bfR_a| \leq R_a$, and tail part, which is in other MT
sites $|\bfr-\bfR_{a'}| \leq R_{a'}$.  As for the tail part, we use the
expansion of \req{onecenter1} as in the case of PW. On the other hand, we
use the head part as it is \cite{privatemark1};
this can be taken into account in the formalism 
if the set $\{\PakL(\bfr) \}$ contains not only the Laguerre
polynomials but also $h_L(\bfr)$ as its members.



After specifying $\{\PakL(\bfr) \}$, we can determine corresponding
$\{\wPakL(\bfr) \}$ as a linear combination of
$\phi_{al}(r)Y_L(\hat{\bfr})$ and $\phidot_{al}(r)Y_L(\hat{\bfr})$,
where partial waves $\phi_{al}(r)$ and its energy derivatives
$\phidot_{al}(r)$ are given as the solutions of the radial Schr\"odinger
equation for the spherically-averaged potential of $V_{1,a}$ in
\req{eq:v}, where energies $E_{al}$ to solve the equation are given as
the center of gravities of the occupied states of the partial density of
states of the $al$ component; thus 
$\phi(r)$ and $\phidot(r)$ are not with the subscripts $aL$ but with $al$.
This prescription to determine $\{\wPakL(\bfr) \}$ can be taken as a
quasi-minimization procedure, from the view of total energy
minimization.

As for the $al$ with Lo, we need another partial wave $\phi^{\rm
Lo}_{al}(r)$ corresponding to Lo.  When the Lo is to describe a deeper
level, we can set the energy to solve the radial Schr\"odinger equation
$E^{\rm Lo}_{al}$ at the center of gravity; then we set $E_{al}$ at the
Fermi energy instead of the prescription in the previous paragraph.

The number of basis is simply specified by the cutoff energy of the APW
for (a). However, specification of MTOs (b) is not so simple.  
We use multiple MTOs for each $aL$ to reduce the number of basis with 
keeping the computational accuracy \cite{pmt1}.
Since $h_L(\bfr)$ as the envelope functions are specified 
by the parameters $\RSM$ and $\epsilon$, we have to specify them for
all MTOs. Ref.\cite{lmfchap} discussed optimization of them so as
to minimize the total energy. However, as seen in figures in
Ref.\cite{lmfchap}, such non-linear optimization is too complicated.
Thus it is necessary to give a method to set the parameters 
in a simple manner as follows. 
As for $\RSM$, we can use a condition $\RSM= R_a/2$ for all MTOs. 
Then the envelope functions out side of MTs well coincide with
the usual Hankel function. 
Even with this simple setting of $\RSM$ without optimization,
numerical accuracy can be kept well; we can check the
convergence of calculations with the number of APWs.
We also see the dependence on $\epsilon$'s are rather small
in the PMT method. The dependence becomes less when we use larger
number of APWs; hence we do not need to stick to careful choice of
the parameter $\epsilon$.
Thus the serious problem of the full-potential LMTO method, ``how
to choose MTO parameters'' are essentially removed in the PMT
method. This is numerically detailed in the paper which gives results for
diatomic molecules from H$_2$ through Kr$_2$ \cite{kotani_linearized_2013}.

We use one further approximation. In \req{eq:n}, we make angular-momentum cutoff.
Even though we have angular momentum component up to 
$2\times l_{{\rm max},a}$ in the 1st and 2nd components in \req{eq:n}, 
we drop components higher than $l_{{\rm max},a}$; 
it is meaningless to take them into account since we have already make 
truncations for eigenfunctions. Note that this does not affects $O_{ij}$
and $T_{ij}$ because only the special components determine them.

\subsection{problems in the PMT method}
\label{sec:problems}
Let us examine three problems of the PMT methods, and ways to
manage them.

The first problem is the positive definiteness of $O_{ij}$.
Since the last term in \req{eq:norm} can give negative contribution,
there is a possibility that $O_{ij}$ can not be positive definite.
In principle, we can expect almost zero eigenvalues on the matrix
$\intaa d^3r \left(F^*_{0i}(\bfr)F_{0j}(\bfr)-F^*_{2i,a}(\bfr)F_{2j,a}(\bfr)\right)$ 
for all MTs if the truncation parameters are large enough. 
This guarantees the positive definiteness of $O_{ij}$.
In practice, we typically use $k_{{\rm max},a} \sim 5$ 
and $l_{{\rm max},a}\sim 4$; they
can give satisfactory results with keeping positive
definiteness of $O_{ij}$, as seen in Refs.\cite{pmt1,kotani_linearized_2013}.

The second is the undefiniteness of the second component $\psi_{2p}$.
This is clear if (A) is satisfied; as $\psi_{2p}$ within MTs is not uniquely
determined since it is canceled completely by $\psi_{0p}$ within MTs.
However, since we use (A') in practice, this can cause numerical instability.
To illustrate this, let us consider a linear combination of basis functions
where only their 0th and 2nd components within MT are non zero.
This is a null vector which has no physical meanings; it gives zero when 
we apply Hamiltonian and Overlap matrix to it. This is a kind of ghost. 
Apparently, this occurs because the 3-component space is not a complete
metric space in the mathematical sense.
When we enlarge number of basis, this null vector can cause numerical problems.
It can be an origin of uncontrollable eigenvalue (e.g, 0 divided by 0), or
it can attach to some eigenfunctions and deform them easily.
In fact, we observed unconverged cases when the 2nd component of
electron density becomes too large. 
Within our current implementation of the PMT, we should use limited number
of basis so as to avoid this problem. However, in Refs.\cite{pmt1,kotani_linearized_2013},
we can see enough stability on the total energy convergence before such problems occurs
when we increase the number of basis .

It will be possible to remove such undefiniteness in some manners.
For example, we can minimize the total energy with adding a 
fixing term $+\lambda \sum_p \intaa d^3r
\psi^*_{2p,a}(\bfr)(1-\tilde{P})\psi_{2p,a}(\bfr)$, 
where $\lambda$ is a Lagrange multiplier, $\tilde{P}$ is a projector to
the space spanned by some pseudo partial waves corresponding to true
atomic partial waves. If $\lambda$ is infinite,
2nd components are only spanned by the pseudo partial waves.
However, we should avoid a large $\lambda$ so as not to 
deteriorate the total energy minimization.

The third problem is the orthogonality to the cores. 
In the frozen core approximation in \refsec{sec:frozencore}, 
we take account of the spillout of the core electron density 
from MTs; this allows us to use a small MT radius.
However, when we use quite small MTs, we observed a problem of orthogonality
of wavefunctions to the cores, resulting in unconvergence. In such a case, 
we  need to introduce local orbitals to represent cores so as to keep
the orthogonality. It may be possible to enforce the orthogonality 
with a projector as described in Ref.\cite{PAW}.

\subsection{comparison with PAW}
\label{sec:comparison}
Here we will make a comparison of the PMT method with the PAW method
\cite{PAW,kresse99} based on the 3-component formalism.

In the PAW method, we perform the all-electron (AE) 
calculations for a spherical atom as a reference in advance.
Then the main problem is how to solve the one-body problem
for a given one-body potential $V(\bfr)$ in real space. 
As in \refsec{aug3}, the problem is translated into the problem in the 3-component space
for $V=V_0\oplus V_1 \ominus V_2$.
For simplicity, we omit the index $a$ in the followings.

The basis set in the PAW is given as follows.
We first prepare AE partial waves
$\{\phi_i(\bfr)\}$ (e.g, two for each $aL$ in Ref.\cite{kresse99}),
as solutions of radial Sch\"odinger eq. for $V_1$ at some reference 
energies $\{\epsilon_i\}$ (in this section, the index $i$ is for the
partial wave). Then we set up corresponding
pseudo partial waves $\{\tilde{\phi}_i(\bfr)\}$.
The eigenfunction $\psi$ in the PAW can be represented 
in the 3-component space; for given 0th-component $\bar{\psi}$
(this is called as {\it pseudo wavefunction}), we have
$\psi$ with projectors $\{\tilde{p}_i\}$ as
\begin{eqnarray}
\psi=\bar{\psi}
\oplus \sum_i |\phi_i\rangle \langle \tilde{p}_i|\bar{\psi}\rangle
\ominus \sum_i |\tilde{\phi}_i \rangle \langle
\tilde{p}_i|\bar{\psi}\rangle.
\label{eq:psibar}
\end{eqnarray}
Here $\tilde{p}_i$ should satisfy
$\langle \tilde{p}_i|\tilde{\phi}_j \rangle=\delta_{ij}$.
The minimization of the total energy of the one-body problem 
$E= \sum_j^{\rm occupied} \psi_j^* \cdot (T+V) \cdot \psi_j$
with respect to $\bar{\psi}_j$ is given by
\begin{widetext}
\begin{eqnarray}
&&\left( \frac{-\nabla^2}{2m}+V_0(\bfr)-\epsilon_j +
 \sum_{ii'}  
  |\tilde{p}_i \rangle  
   \left(dH_{ii'}-\epsilon_j dO_{ii'} \right)  
  \langle \tilde{p}_{i'}| 
\right) \bar{\psi}_j=0, \label{eq:seqpsibar}\\
&&dH_{ii'}= \langle \phi_i| \frac{-\nabla^2}{2m}+V_1|\phi_{i'}\rangle  
 -  \langle \tphi_i| \frac{-\nabla^2}{2m}+V_2|\tphi_{i'}\rangle \\
&&dO_{ii'}= \langle \phi_i|\phi_{i'}\rangle  
 -  \langle \tphi_i| \tphi_{i'}\rangle.
\end{eqnarray}
If we use infinite number of partial waves which makes a complete set, 
\req{eq:seqpsibar} reproduces the original one-body problem in real space.

Let us consider a case where
$\psi_j=\bar{\psi}_j \oplus \phi_j\ominus\tphi_j$
is the solution of \req{eq:seqpsibar} with eigenvalue $\epsilon_j$,
where $\bar{\psi}_j$ within MT coincides with $\tphi_j$.
This is given by \req{eq:psibar} from $\bar{\psi}_j$.
When we make a truncation for the number of partial waves,
$\{\tilde{p}_i\}$ should satisfy 
\begin{eqnarray}
&& \left( \frac{-\nabla^2}{2m}+V_0(\bfr)-\epsilon_j \right)|\tphi_j \rangle
 + \sum_i |\tilde{p}_i \rangle  
   \left(dH_{ij}-\epsilon_j dO_{ij} \right)  =0, \label{eq:condproj}
\end{eqnarray}
\end{widetext}
in order to satisfy \req{eq:seqpsibar}.
This determines $\{\tilde{p}_i\}$; this is one of the main idea in the PAW method. 
In practice, considering the numerical stability, 
we determine  $\tilde{p}_i$ so that \req{eq:condproj} is approximately
satisfied \cite{PAW}. 

Another important idea of the PAW is the introduction of the
pseudopotential. This is how to determine $V_0$ within MT ($=V_2$).
This is because the result strongly depends on the pseudopotential
when the number of partial waves are small. In principle, the pseudopotential 
should be determined so that $\bar{\psi}_j$ contain high energy part 
(high angular momentum $l$ or highly oscillating part) of the wavefunctions
which is missing in the 1st and 2nd components due to the 
truncation of the number of partial waves.

Note that the truncation can cause the ghost state problem in the PAW method.
To illustrate this, consider a case that $s$ wave part in MT is described
only by two partial waves $2s$ and $3s$. Then the PAW procedure maps $\bar{\psi}$ with zero node to 
$\psi$ with one node, $\bar{\psi}$ with one node to $\psi$ with two nodes. 
Problem is that $\bar{\psi}$ with two nodes, which is orthogonal to $\{\bar{\psi}_i\}$ for $2s$ and $3s$,
can not be mapped to $\psi$ with three nodes due to the truncation. 
Thus it is possible that such a function cause a ghost state;
we have to design the pseudopotential so that such $\bar{\psi}$ should be
kept to be at a high enough energy region (to push $\bar{\psi}$ high away from the
Fermi energy, it may be better to use relatively strongly repulsive pseudopotential). 
Ref.\cite{kresse99} claims that there is no
ghost state for all kinds of atoms. However, it is not easy
to check the convergence within the framework of the PAW method.

In the PAW method with PWs proposed in Ref.\cite{kresse99}, many PWs are required
compared with with LAPW. Roughly speaking, energy cutoff of PWs are
$\sim$15Ry in LAPW, and $\sim$30Ry in PAW \cite{filippi94,kresse99}.
This is because the PAW method, as is the case of pseudopotential methods,
needs to uniquely determine the pseudo partial waves (0th component) within MT.
This is in contrast with the LAPW (and the PMT) method, 
where 0th component within MT is irrelevant because 
the 2nd components have enough degree of freedom to well cancel its contribution.
However, with sacrificing the cutoff energy,
the PAW takes robust convergence that comes from the absence of the 
null vector problem discussed in \refsec{sec:problems}

As a theoretical possibility, we can imagine a method to 
use \smhs\ together with the PWs in the basis set
for the one-body problem in the PAW method.
However, it is not very clear whether it becomes 
a efficient method or not. To reduce the number of basis of PWs, it is
necessary to make the \smhs\ span high-energy parts of pseudo
wavefunctions. Thus we have to tailor \smh\ so that it fits to the pseudo
wavefunctions not only interstitial region, but also within MT. 
This can be not straightforward.

\section{summary}
We have reformulated the PMT method on the basis of the 3-component
formalism, which is a generalized version of the additive augmentation 
given by Soler and Williams. The 3-component formalism
allows including any kinds of basis not necessarily given by a
projector as PAW. This fits the procedure to give the Kohn-Sham equation for a mixed basis method such 
as the PMT method from the total energy minimization scheme; this
results in the transparent derivation of the atomic forces.
We believe that the formalism shown here could give a basis for future developments.
Our results for molecules from H$_2$ through Kr$_2$ 
with several new developments on the PMT method 
is given elsewhere \cite{kotani_linearized_2013}.

\  \\
Acknowledgement:
We appreciate discussions with Profs. T.Oguchi, S.Bluegel, and P.Blochel.
This work was supported by Grant-in-Aid for Scientific Research 23104510.

\appendix
\begin{widetext}
\section{the error due to the separable form}
\label{sec:zeroonetwo}
To evaluate matrix element of a quasilocal operator
$X(\bfr,\bfr')$, we use separable form $0X 0'+1 X 1'-2 X 2'$
instead of $(0+1-2)X(0'+1'-2')$ under the condition (A') 
(see \refsec{sec:3compo}). Here $0,1,2$ means the three components of 
a eigenfunction as a 3-component function defined in
\refsec{sec:3compo}, $0',1',2'$ as well.

We have an error because of the separable form.
Here we reorganize the discussion to evaluate the error
\cite{soler89,PAW} to fit to the formalism in this paper.
The error can be evaluated with an identity as;
\begin{eqnarray}
(0+1-2)X(0'+1'-2')-(0X 0'+1 X 1'-2 X 2') 
= (0-2)X(1'-2')+(1-2)X(0'-2'), \label{eq:zeroonetwo}
\end{eqnarray}
\end{widetext}
Let us examine the error as the right-hand side of
\req{eq:zeroonetwo} under the assumption that $X$ is nearly spherical.
Remember that $(0-2)$ is completely zero if the condition (A) is
satisfied. When the condition (A') is satisfied instead, 
i.e., when we introduce the finite truncation parameters 
$l_{{\rm max},a}$ and $k_{{\rm max},a}$ (given after \req{f2}), 
we can expect that $(0-2)$ should contain high-energy remnant
(high angular momentum $l$ or highly oscillating remnant) with a small amplitude.
The remnant $(0-2)$ for each $L$ is largest at the MT boundaries.
In contrast, when (A') is satisfied, $(1'-2')$ is low energy part which  
converges quickly on the truncation parameters. 
The value and slope of $(1'-2')$ are zero at MT boundaries.
Thus we can expect the product $(0-2)(1'-2')$ should be small and
nearly orthogonal, i.e.,
$\delta n_a(\bfr)=(0-2)_a(1'-2')_a$ should satisfy $\int_a d^3 r\delta
n_a(\bfr)Y_L(\hat{\bfr}) \approx 0$ for low $L$.
Here suffix $a$ means quantities within MT at $\bfR_a$.
Based on these considerations we expect that the error affects little 
the total energy.
This can be checked by changing the truncation parameters within the PMT method.

This logic is applicable not only to the products of the eigenfunctions, 
but also to the electron density for the Coulomb interaction 
with some modifications.



\begin{widetext}
\section{Atomic Force}
\label{sec:force}
First, we define the Harris energy $\ehf$ \cite{molforce,harris85} 
which is the total energy of a functional of the density; this gives
a reasonable estimate of the total energy even when the density is
somehow different from the converged density.
When not being converged yet, the input density $\nin$
must be treated as one generating the one-particle potential $V$,
and output density $\nout$ which is given from eigenfunctions
obtained from the eigenvalue problem of $V$. Here, $
V$ is given by \req{eq:v}. Now, $\ehf$ in the frozen core
approximation as a functional of $\nin$ is defined by \cite{molforce}:
\begin{eqnarray}
&&\ehf = E_{\rm k}^{\rm core} + E_{\rm B} - V[\nzc+\nin,\bfR_a] \cdot \nin 
+ E_{\rm es}[\nzc+\nin,\bfR_a] + E_{\rm xc}[\nzc+\nin],
\label{eq:ehf} \\
&&E_{\rm B} = \sum_p^{\rm occupied}
\alpha_{p}^{i*} 
\langle F_i|H^{\rm in}|F_j \rangle 
\alpha_p^j,
\label{eq:ebhf}
\end{eqnarray}
where $E_{\rm B}$ is the band energy.
$\alpha_i^p$ is the eigenvector of 
$\langle F_i|H^{\rm in}|F_j \rangle=
\langle F_i| \frac{-\Delta}{2m} + V[\nzc+\nin,\bfR_a] |F_j\rangle$.
Thus we have $E_{\rm B}=\sum^{\rm occupied}_p \epsilon_p$, where $\epsilon_p$ are eigenvalues.
The $\bfR_a$-dependence explicitly shown in \req{eq:ehf} is
through the $\MM$-transformation and $\calR$-mapping;
even when $\nzc+\nin$ is fixed as a 3-component function,
$\bfR_a$-dependence is introduced to \req{eq:v} through Eqs.(\ref{eq:barn0zcv},\ref{eq:calqdef}).
In addition, we have $\bfR_a$-dependence through $\nzc+\nin$.

Atomic forces are given as the change of the total energy 
for atomic displacement $\delta \bfR_a$.
Here let us consider the change of $\ehf$, written as $\delta \ehf$.
To obtain $\delta \ehf$, we use the derivative chain rule where we
treat $\ehf$ as a function of $\bfR_a$ through
$\{F_i(\bfr), \nin, \Vin, \bfR_a\}$; $\Vin$ means $V[\nzc+\nin,\bfR_a]$ in 
Eqs.(\ref{eq:ehf},\ref{eq:ebhf}).
Remember that there is $\bfR_a$ dependence through $\nzc$.
Here we assume the partial waves
($\{\phi_{al}(r),\dot{\phi}_{al}(r),\philo(r) \}$ in the case of the PMT
method) are not dependent on atomic positions as in Ref.\cite{molforce}.

Let us evaluate $\delta \ehf$. 
As for $E_{\rm B}=\sum_p^{\rm occupied} \epsilon_p$ 
as a functional of $\{F_i(\bfr),\Vin\}$, perturbation theory on
\req{eq:eigenp} gives
\begin{eqnarray}
&&\delta E_{\rm B} = \sum_p^{\rm occupied} \delta \epsilon_p
=\sum_p^{\rm occupied} \sum_i \sum_j 
\alpha_{p*}^i (\delta H_{ij} -\epsilon_p \delta O_{ij}) \alpha_p^j
=\delta \Vin\cdot \nout + \delta E_{\rm B}^{\rm Puley}, \label{eq:deleb}\\
&&\delta E_{\rm B}^{\rm Puley}=\sum_p^{\rm occupied} \sum_i \sum_j 
\alpha_{p}^{i*} (\delta H^F_{ij} -\epsilon_p \delta O^F_{ij}) \alpha_p^j,
\end{eqnarray}
where we have used $\delta (\Vin\cdot F^*_i F_j)=\delta \Vin 
\cdot F^*_i F_j + \Vin\cdot \delta (F^*_i F_j)$.
$\delta E_{\rm B}^{\rm Puley}$ is calculated from 
$\delta F_{0i}(\bfr)$ and $\delta C^i_{akL}$, which 
are given as a functional of $\delta \bfR_a$.

Since $E_{\rm es} + E_{\rm xc}$ is a functional of $\{\nin,\bfR_a\}$, we have
\begin{eqnarray}
&&\delta \ehf = \delta E_{\rm B} - \delta (\Vin \cdot \nin) 
  + \delta (E_{\rm es} + E_{\rm xc}) \nonumber \\
&&= \delta \Vin \cdot (\nout- \nin) 
  + \delta E_{\rm B}^{\rm Puley} 
  + \left.\frac{\partial (E_{\rm es} + E_{\rm xc})}{\partial \bfR_a}\right|_{\nin} \delta \bfR_a.
\label{eq:deltaehf} 
\end{eqnarray}
There are three terms in the right hand side of \req{eq:deltaehf}.
The first term appears because $\ehf$ is not converged yet.

To calculate the first term, we need to know $\delta \nin$ which determines
$\delta \Vin$. When the self-consistency is attained and converged,
that is, $\nin=\nout$, $\delta \bfR_a$ uniquely determines
$\delta \nin=\delta \nout$.
However, this is not true when $\nin \ne \nout$.
In this case, there is no unique way to determine $\delta \nin$
for given $\delta \bfR_a$. Thus we need an extra assumption to
determine it. As a reasonable and convenient choice,
we use $\delta \nin=0$ in the sense of 3-component representation. 
Physically, this means that $n_{1,a}(\bfr)-n_{2,a}(\bfr)$  
together with frozen core centered at $\bfR_a$
moves rigidly to $\bfR_a+\delta\bfR_a$.
Then we can calculate corresponding $\delta \Vin$ through the change
$\delta \barnzcv_0$ in \req{eq:v}. $\delta \barnzcv_0$ is evaluated from
\req{eq:barn0zcv},
where note that $\nzc_0(\bfr)$ contains $\bfR_a$ dependence as
given in \req{eq:nzc}.

\section{onsite matrix}
\label{onsitematrix}
Here we summarize expressions of one-center matrix for $O_{ij},T_{ij}$, and
$V_{ij}$. These are essentially the same as what is shown in Ref.\cite{lmfchap}.
With the help of Eqs.(\ref{f2},\ref{f1}), Eqs(\ref{eq:norm},\ref{eq:kin},\ref{eq:v})
are reduced to be
\begin{eqnarray}
O_{ij} &=&\int_\Omega d^3r  F^*_{0i}(\bfr)F_{0j}(\bfr)
  + \sum_{akk'L} C^{*i}_{akL} \sigma_{akk'L} C^{j}_{akL}   \label{eq:normmat1}\\
T_{ij} &=&\frac{1}{2m_e}\int_\Omega d^3r  \nabla F^*_{0i}(\bfr) \nabla F_{0j}(\bfr)
  + \sum_{akk'L} C^{*i}_{akL} \tau_{akk'L} C^{j}_{ak'L},   \label{eq:kinmmat1}\\
V_{ij}&=&\int_\Omega d^3r  F^*_{0i}(\bfr)V_0(\bfr)F_{0j}(\bfr)
  + \sum_{akk'LL'} C^{*i}_{akL} \pi_{akk'LL'} C^{j}_{ak'L'} \label{eq:vpot},
where
\end{eqnarray}
\begin{eqnarray}
&&\sigma_{akk'l}= \inta d^3r  
 \left(\widetilde{P}_{akL}(\bfr) \widetilde{P}_{ak'L}(\bfr)
- {P}_{akL}(\bfr) {P}_{ak'L}(\bfr)\right), \label{matsig} \\
&&\tau_{akk'l}= \frac{1}{2m_e}\inta d^3r  
 \left(\nabla \widetilde{P}_{akL}(\bfr) \nabla \widetilde{P}_{ak'L}(\bfr)
-\nabla {P}_{akL}(\bfr) \nabla {Pq}_{ak'L}(\bfr)\right), \label{mattau}\\
&&\pi_{akk'LL'}= \sum_M Q_{kk'LL'M} {\cal Q}^{\rm v}_{aM} +
  \left(\barnzcv_{1,a} \circ \RR + v^{\rm xc}_{1,a} \right)\circ
  \widetilde{P}_{akL}(\bfr') \widetilde{P}_{ak'L}(\bfr')
- \left(\barnzcv_{2,a} \circ \RR + v^{\rm xc}_{2,a} \right) \nonumber \\
&&\circ
  {P}_{akL}(\bfr') {P}_{ak'L}(\bfr'), \label{matpi}\\
&&Q_{kk'LL'M}=\inta d^3r
\left( \widetilde{P}_{akL}(\bfr) \widetilde{P}_{ak'Ll}(\bfr)
- {P}_{akL}(\bfr) {P}_{ak'Ll}(\bfr)\right) \YY_M(\bfr) \label{qmom}. 
\end{eqnarray}
\end{widetext}
Note that $\sigma_{akk'l}$ and $\tau_{akk'l}$ are dependent only on
$l$ of $L=(l,m)$.
In Ref.\cite{lmfchap}, this $\pi_{akk'LL'}$ is further divided as
$\pi^{\rm mesh}_{akk'LL'} + \pi^{\rm local}_{akk'LL'}$.
${\cal Q}^{\rm v}_{aM}$ is given by \req{eq:calqdef}.

\section{scalar relativistic approximation in the augmentation}
\label{app:srel}
Roughly speaking, it is allowed to take the scalar relativistic
(SR) approximation (e.g. see \cite{rmartinbook}) if we can safely replace the
non-relativistic (NR) wavefunctions with the SR wavefunctions within MTs.
The SR wavefunctions contain major and minor components. 
The major component should be smoothly connected to the NR wavefunction in 
the interstitial region, where the minority parts are negligible. 
All physical quantities within MT should be evaluated 
through the SR wavefunctions.
In the followings, we explain how the above idea can be implemented in the
3-component augmentation for bilinear products. 

First, we modify the 1st component of the basis. 
We use two component wavefunctions $\{{\mathtt g}_{1i,aL}(\bfr),{\mathtt f}_{1i,aL}(\bfr)\}$
instead of $F_{1i,a}(\bfr)$, where the SR approximation gives
${\mathtt f}_{1i,a}(\bfr)=\frac{1}{2m_e c} \frac{d {\mathtt g}_{1i,a}(\bfr)}{dr}$, 
where $c$ is the speed of light. 
For given $F_{0i}$ and $F_{2i}$ (they are the same as those of the NR case),
we ask the the major components ${\mathtt g}_{1i,a}(\bfr)$
to satisfy the boundary conditions as for value and slope at MT boundaries.

In order to calculate the contributions due to the 1st components within the SR approximation
instead of the NR approximation, we make a replacement 
$F^*_{1i,a}(\bfr)F_{1j,a}(\bfr') \rightarrow 
 {\mathtt g}^*_{1i,a}(\bfr){\mathtt g}_{1j,a}(\bfr')
+\left(\frac{1}{2m_e c}\right)^2 
 {\mathtt f}^*_{1i,a}(\bfr){\mathtt f}_{1j,a}(\bfr')$.
With this replacement, we can evaluate the density $n$, the matrix $O_{ij}$ 
and so on. This ends up with the total energy in the SR approximation.

Finally, we see that changes are in the replacement Eqs.(\ref{matsig}-\ref{qmom}),
where products $\widetilde{P}_{akL}(\bfr) \widetilde{P}_{ak'L}(\bfr)$ 
(and those with $\nabla$) should be interpreted not only from the products 
of the majority wavefunctions, but also from those of the minority.
This occurs also for the density $n_{1,a}$ included in \req{matpi}.

In such a way we can include the SR effect in the 3-component formalism.
In a similar manner, we can include the spin-orbit coupling in the 1st component, 
which results in the spin off-diagonal contributions \cite{chantis06a}.

\bibliography{lmto,gw}

\begin{thebibliography}{31}
\bibitem [{\citenamefont {van Schilfgaarde}()}]{privatemark1}%
  \BibitemOpen
  \bibfield  {author} {\bibinfo {author} {\bibfnamefont {M.}~\bibnamefont {van
  Schilfgaarde}},\ }\href@noop {} {}\bibinfo {note} {Private communication: the
  head part of the smooth Hankel functions are difficult to be expanded in the
  series of polynomials accurately.}\BibitemShut {Stop}%
\end{thebibliography}
\end{document}